\documentclass{elsart}
\usepackage{amssymb}
\usepackage{graphicx}
\usepackage{psfrag}

\begin{document}
\runauthor{Guimer\`a, Arenas, D\'{\i}az}
\begin{frontmatter}
\title{Communication and optimal hierarchical networks\thanksref{X}}
\author[DEQ]{R. Guimer\`a}
\author[DEIM]{A. Arenas}
\author[UB]{A. D\'\i az-Guilera}
\thanks[X]{Partially supported by DGES of the Spanish Government,
grants PB96-0168 and PB97-0131, and EU TMR grant ERBFMRXCT980183}

\address[DEQ]{Departament d'Enginyeria Qu\'{\i}mica, Universitat Rovira i Virgili, Carretera Salou s/n, E-43006 Tarragona, Spain}
\address[DEIM]{Departament d'Enginyeria Inform\`atica, Universitat Rovira i Virgili, Carretera Salou s/n, E-43006 Tarragona, Spain}
\address[UB]{Departament de F\'\i sica Fonamental, Universitat de Barcelona, Diagonal 647, E-08028 Barcelona, Spain}

\begin{abstract}
We study a general and simple model for communication processes. In the model, agents in a network (in particular, an organization) interchange information packets following simple rules that take into account the limited capability of the agents to deal with packets and the cost associated to the existence of open communication channels. Due to the limitation in the capability, the network collapses under certain conditions. We focus on when the collapse occurs for hierarchical networks and also on the influence of the flatness or steepness of the structure. We find that the need for hierarchy is related to the existence of costly connections.
\end{abstract}
\end{frontmatter}

Nowadays, physicists pay a lot of attention to the dynamics of complex social and economic systems \cite{arthur97,axelrod97,axelrod99}. In particular the influence of the topology of the underlying interactions on the behavior of such systems \cite{watts98,barabasi99,amaral00} deserves special interest. We focus on the behavior of hierarchical structures formed by agents (or elements, in general) that interact with each other via communication processes. This framework is especially adequate to study for instance packet flow in computer networks like the Internet \cite{takayasu96,ohira98}, traffic networks \cite{chowdhury00}, river networks \cite{banavar99} and particularly information flows in organizations \cite{radner93,vanzandt99,garicano00}.

Using Radner's words \cite{radner93}:
\begin{quote}The typical U.S. company is so large that a substantial part of its workforce is devoted to information-processing, rather than to ``making'' or ``selling'' things in the narrow sense. Although precise definitions and data are not available, a reasonable estimate is that more than one-half of U.S. workers (including managers) do information-processing as their primary activity.
\end{quote}
Thus it is worth considering an organization as a system of information processors.

In this work, we extend a general and simple model for communication processes that has been recently proposed \cite{arenas01}. The original model considers agents that deliver information packets through well established channels. These agents have an infinite capacity to store packets and the only limitation comes from the fact that they do not have an infinite capacity to deliver. Despite its simplicity, the model reproduces the main characteristics of the flow of information packets in a network and a continuous phase transition from a free to a congested regime is observed and properly characterized by means of an order parameter \cite{arenas01}.

In the present extension of the model, we introduce a cost associated to the establishment of links so that the agents in the communication network cannot be linked to an arbitrary number of neighbours or, at least, it has a negative influence on their performance. In both cases, with and without cost associated to the links, the optimal organizational structures are studied.

\begin{figure}[h]
\vskip 0.2in
\centerline{\includegraphics*[width=0.6\columnwidth]{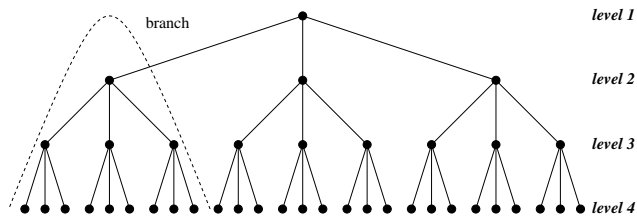}}
\caption{Typical hierarchical tree structure used for simulations and calculations: in particular, it is a tree $(3,4)$. Dashed line: definition of {\it branch}.
\label{network}}
\end{figure}

The organization is mapped into a hierarchical (Bethe) lattice (see Fig.~\ref{network}) where nodes represent the communicating agents (employees) and the links between them represent communication lines. These tree like structures are characterized by two quantities: the branching factor, $z$, and the number of levels, $m$. From now on, we will use the notation $(z,m)$ to describe a particular tree.

The dynamics of the model is the following. At each time step $t$, an information packet is created by every agent with probability $p$. When a new packet is created, a destination agent, different from the origin, is chosen at random in the network. Thus, during the following time steps $t, t+1,\dots ,t+T$, the packet is traveling toward its destination: once the packet  reaches this destination agent, it is delivered and disappears from the network. One can think in a problem solving scenario \cite{garicano00} and say that, from time to time, problems arise in the organization; these problems need to be solved somewhere in the network. When an agent receives a packet (problem), she knows whether the destination (solution) is to be found somewhere below her. If so, she directs the packet downwards in the right direction. Otherwise, she transmits it upwards to the agent overseeing her. Thus, the information packets move toward their destination following the shortest path. The time a packet remains in the network is related not only to the distance between the origin and the destination agents, but also to the amount of packets in the network. In particular, at each time step, all the agents try to send each one of the packets they are handling. For each packet, there is a probability $q_{ij}$ to go from the present agent $i$ to the next one $j$. We call $q_{ij}$ the {\it quality of communication} between agents $i$ and $j$ and it is defined as
\begin{equation}
q_{ij}=\sqrt{k_{i}k_{j}}.
\end{equation}
where $k_{\alpha}$ represents the capability of agent $\alpha$ to communicate at each time step. For $k_{\alpha}$ we propose:
\begin{equation}
k_{\alpha}=Q_L(c_\alpha)f(n_\alpha)
\end{equation}
where $c_\alpha$ is the number of links of agent $\alpha$, $0<Q_L(c)\le1$ is a cost factor related to these links (note that, the higher the number of links, the smaller $Q_L$, so $Q_L$ is a monotonically decreasing function of its argument), $L$ is the {\it linking capability} that tunes the magnitude of this cost (higher values of $L$ correspond to low linking cost and viceversa), $n_\alpha$ is the total number of packets currently at agent $\alpha$, and $0<f(n)\le1$ is the function that determines how the capability of a particular agent decreases when the number of information packets to handle grows (again, $f(n)$ is a decreasing function of the argument).

A suitable election for $f(n)$ seems to be
\begin{equation}
f(n)=\left\{ \begin{array}{lcl}
	1	&	\quad\mbox{for}	&	n=0\\
	1/n 	&	\quad\mbox{for}	&	n=1,2,3,\dots
	\end{array}\right.
\end{equation}
although other functional forms can be considered and one observes different interesting behaviours \cite{arenas01,guimera??}.

As a first step, let us focus in the simpler case $L\rightarrow\infty$, i.e. cost less connections. The probability of generating a packet per agent and time unit, $p$, is an exogenous parameter that controls the behavior of the system. For small values of $p$, all the packets are delivered and so, after a transient, the system reaches a steady state in which the total number of packets, $N$, fluctuates around a constant value, i.e. the number of delivered packets is equal, on average, to the number of generated packets. However, for large values of $p$, not all the packets can be delivered, and $N$ grows in time without limit. The transition between one regime  and the other is a continuous phase transition and occurs for a well defined critical value of $p$, $p_c$ \cite{arenas01}.

It is possible to give an analytical estimation of $p_c$. Within a mean field approach, it is if we do not consider fluctuations and we assume that the behavior of all the agents in the same level is statistically identical, one gets the following expression for $p_c$ \cite{arenas01}
\begin{equation}
p_c=\frac{\sqrt{z}}{\frac{z(z^{m-1}-1)^2}{z^m-1}+1}.
\label{p_c_calc}
\end{equation}
For values of $z$ and $m$ such that $z^{m-1}\gg1$ (note that this condition is satisfied even for relatively small values of $z$ and $m$), this expression can be approximated by
\begin{equation}
p_c\approx z^{3/2-m}.
\label{p_c_2_calc}
\end{equation}
Although strictly speaking (\ref{p_c_calc}) (and its approximation (\ref{p_c_2_calc})) provides an upper bound to $p_c$, it is an excellent estimation for $z\ge4$, as can be seen in Fig.~2 of Ref.~\cite{arenas01}.

More interesting to us is the maximum number of information packets that can be generated in a time step without collapsing the organization, $N_c=p_cS$, with $S$ standing for the size of the organization. It is given by
\begin{equation}
N_c=\frac{\sqrt{z}}{\frac{z(z^{m-1}-1)^2}{z^m-1}+1}\frac{z^m-1}{z-1}\approx\frac{z^{3/2}}{z-1}
\end{equation}
again with the same approximation as in (\ref{p_c_2_calc}). Thus the total number of packets a network can deal with does not depend on the number of hierarchical levels.  Furthermore $N_c$ is a monotonically increasing function of $z$, suggesting that, fixed the number of agents in the organization, $S$, the optimal organizational structure, understood as the structure with higher capacity to handle information, is the flattest one, with $m=2$ and $z=S-1$.

However, from a practical point of view this structure is not possible: an organization with 10,000 employees, for instance, cannot be organized in only two hierarchical levels, since it is impossible to maintain such a enormous number of communication lines. Thus, it is necessary to introduce the cost for establishing links in order to get a more realistic picture of the problem. In this case, following arguments analogous to that used in the case of cost less connections, we can arrive to the following expression for $p_c$:
\begin{equation}
p_c=\frac{\sqrt{zQ_L(z)Q_L(z+1)}}{\frac{z(z^{m-1}-1)^2}{z^m-1}+1}.
\end{equation}

Again, for $z$ and $m$ such that $z^{m-1}\gg1$, the maximum number of packets that can be generated per time step without collapsing the system is independent of $m$, and is given by
\begin{equation}
N_c\approx\frac{z^{3/2}(Q_L(z)Q_L(z-1))^{1/2}}{z-1}.
\end{equation}

To check the effect of the cost factor, we propose the following form for $Q_L(c)$
\begin{equation}
Q_L(c)=1-\tanh{\frac{c}{L}},
\label{C}
\end{equation}
Although the election of $Q_L$ is completely arbitrary, (\ref{C}) has two desirable properties: (i) it is a monotonically decreasing strictly positive function and (ii) $Q_L$ decreases linearly for small values of $c$ (compared to $L$). Also, $Q_L$ decreases faster for small values of $L$ and viceversa.

\begin{figure}[t]
\centerline{\includegraphics*[width=0.6\columnwidth]{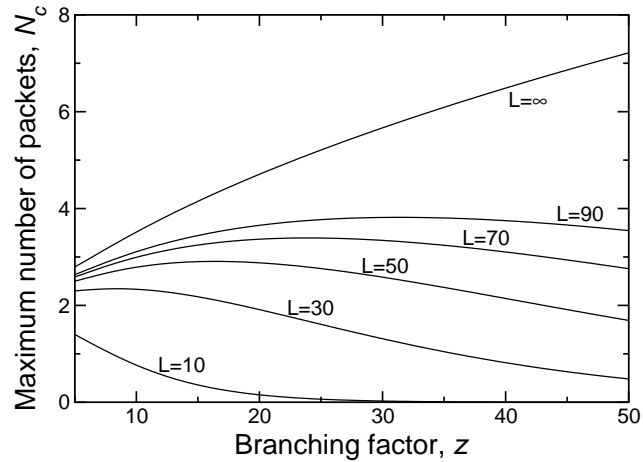}}
\caption{Maximum number of packets that can be generated in an organization per time unit without collapsing it, plotted as a function of $z$. Different curves correspond to different values of the linking capability, $L$.
\label{cost_N_c}}
\end{figure}

As can be seen from Fig.~\ref{cost_N_c}, the scenario that arises with the introduction of the cost factor is much more interesting. Now, the cost term compete with the behavior we have found for the critical number of generated packets, $N_c$, in the case of cost less connections. Thus, there is a maximum in $N_c$ related to an optimum value of $z$, $z^*$, which defines an optimal organizational structure different from the trivial $m=2$ and $z=S-1$.

Summarizing we have studied a model of communication that includes cost for establishing communication channels. While in the absence of such cost the flat structure is the most efficient (in spite of the congestion phenomena), the need for hierarchy arises because of the existence of even small costs. Remarkably, in both cases with and without costs, the capacity of the network to handle information packets does not depend on the number of levels of the hierarchy but in the branching factor of the structure.


\end{document}